\documentclass[aps,prl,twocolumn,superscriptaddress]{revtex4-1}

\usepackage[dvips]{graphicx}
\usepackage{tabularx}
\usepackage{subfigure}
\usepackage{amsmath}
\usepackage{amsmath}
\usepackage{amstext}
\usepackage{amssymb}
\usepackage{color} 

\begin{document}


\title{Fluctuations in Hertz chains at equilibrium}


\author{Michelle Przedborski}
\affiliation{Department of Physics, Brock University, St. Catharines, Ontario, Canada L2S 3A1}

\author{Surajit Sen}
\affiliation{Department of Physics, State University of New York, Buffalo, New York 14260-1500, USA}

\author{Thad A. Harroun}
\affiliation{Department of Physics, Brock University, St. Catharines, Ontario, Canada L2S 3A1}


\date{\today}

\begin{abstract}
We examine the long-term behaviour of non-integrable, energy-conserved, 1D systems of macroscopic grains interacting via a contact-only generalized Hertz potential and held between stationary walls. Existing dynamical studies showed the absence of energy equipartitioning in such systems, hence their long-term dynamics was described as \emph{quasi-equilibrium}.
Here we show that these systems do in fact reach thermal equilibrium at sufficiently long times, as indicated by the calculated heat capacity. As a byproduct, we show how fluctuations of system quantities, and thus the distribution functions, are influenced by the Hertz potential. In particular, the variance of the system's kinetic energy probability density function is reduced by a factor related to the contact potential. 
\end{abstract}


\maketitle


Recently, there has been broad interest in 1D systems of macroscopic grains held between stationary walls and interacting via a power-law contact potential~\cite{Nesterenko1983,*Nesterenko1985,*Nesterenko1995,*Sinkovits1995,*Sen1996,*Coste1997,*Sen1998,*Chatterjee1999,*Hinch1999,*Hong1999,*Ji1999,*Manciu1999,*Manciu1999a,*Hascoet2000,*Sen2001,*Sen2001b,*Nesterenko2001,*Rosas2003,*Rosas2004,*Herbold2009,*Vitelli2012,Sen2008,Sen2004,*Mohan2005,*Avalos2011,Sen2005,*Avalos2007,*Avalos2014,Manciu2000,*Manciu2002,*Job2005,*Nesterenko2005,*Daraio2006,*Job2007,*Sokolow2007,*ZhenYing2007,*Santibanez2011,*Takato2012,Przedborski2015,*Przedborski2015b}. A long-standing open problem is whether thermalization (equipartition) can occur in these chains of grains. Only very recently has it been shown that the related FPU chain of coupled oscillators does reach equilibrium after very long times~\cite{Onorato2015}. In this paper, we show this is also true for so-called Hertz chains. In the process, we obtain wholly new approximate distribution functions for \emph{interacting} particles in the  \emph{microcanonical} ensemble.

Many power-law interacting systems are notable for supporting solitary wave (SW) propagation~\cite{Przedborski2015,Mohan2005,Valkering1980}. However, in response to singular perturbations, the breakup of SWs at the walls and from gaps between grains leads the system after a long time to an equilibrium-like, ergodic phase~\cite{Sen2008,Sen2004,*Mohan2005,*Avalos2011,Sen2005,*Avalos2007,*Avalos2014}. Unusually large~\cite{Sen2008,Sen2004,*Mohan2005,*Avalos2011,Sen2005,*Avalos2007,*Avalos2014} and occasionally persistent (rogue)~\cite{Han2014} fluctuations in the system's kinetic energy are seen at late times for sufficiently strong and unique perturbations. This has been seen to impede an equal sharing
of energy among all the grains in the system, hence the long-term dynamics of 1D systems of interacting grains has been described as quasi-equilibrium (QEQ) ~\cite{Sen2008,Sen2004,*Mohan2005,*Avalos2011,Sen2005,*Avalos2007,*Avalos2014}. The question of whether QEQ is the final state for these systems is addressed in this letter. 


To the time scales previously studied, quasi-equilibrium has been seen  to be a general feature of the dynamics of systems with no sound propagation ~\cite{Sen2004,*Mohan2005,*Avalos2011}. However, we find that at sufficiently late times, kinetic energy fluctuations relax, allowing for energy to be shared equally among all grains. Of course, energy equipartitioning happens only in an average sense in finite systems, and at any given instant each grain will not have exactly the same kinetic energy. Rather, each grain's kinetic energy fluctuates according to the same probability density function (pdf), the long tail of which determines the chance of large fluctuations. 

The fluctuations are quantified by treating the chain as a 1D gas of interacting spheres~\cite{Scalas2015}. This requires new velocity and kinetic energy distribution functions different from hard spheres, which incorporate the interaction potential. These distributions are also influenced by the finite heat capacity of the system,  which governs the fluctuations in the system kinetic energy in a microcanonical ensemble~\cite{Lebowitz1967}. An equilibrium value for the specific heat obtained using Tolman's generalized equipartition theorem~\cite{Tolman1918}, provides a direct way to probe the extent to which energy equipartitioning occurs in large but finite systems. We show that at sufficiently long times, calculated specific heat capacities of chains of interacting grains agree with the values predicted by the generalized equipartition theorem, indicating that energy equipartitioning holds, and consequently that the ultimate fate of these systems is a true equilibrium phase that can be described by statistical mechanics.

The specific systems under consideration are 1D chains of $N$ grains, each with mass $m$ and radius $R$, interacting via a Hertz-like contact-only potential~\cite{Hertz1882}. The Hamiltonian describing the system is:
\begin{equation}
\mathbf{H} = K+U = \frac{1}{2} \sum_{i=1}^N m v_i^2 + \sum_{i=1}^{N-1} a \Delta_{i,i+1}^n,
\label{eq:hamiltonian}
\end{equation}
where $v_i$ is the velocity of grain $i$ and $\Delta_{i,i+1} \equiv 2R - (x_{i+1}-x_i) \ge 0$ is the overlap between neighbouring grains, located at $x_i$. If $\Delta_{i,i+1}<0$, there is no potential interaction. In the above expression, the exponent $n$ is shape dependant ($n=2.5$ for spheres), and $a$ contains the material properties of the grains~\cite{Sun2011}. The grain interactions with the fixed walls adds two terms to the Hamiltonian, cf. Ref.~\cite{Przedborski2015,*Przedborski2015b}. 

The pdf of particle velocity of a $d$-dimensional, finite sized microcanonical ensemble is not a Maxwell-Boltzmann distribution~\cite{Ray1991, Scalas2015}. The actual distribution can be found from the total volume of a $2dN$-dimensional phase space circumscribed by the total energy $E$,
\begin{equation}
\Omega \propto \int \Theta\left(E-\mathbf{H}\right) dq^{dN} dp^{dN},
\label{eq:ps_vol}
\end{equation}
where $\Theta$ is the Heaviside step function. The integral in Eq.~(\ref{eq:ps_vol}) is taken over all grain momenta $\bf{p}$ and all grain positions $\bf{q}$. Integration over the grain momenta evaluates to the volume of a $dN$-dimensional hypersphere of radius $[2m(E-U)]^{1/2}$, leaving the remaining integral over the grain positions:
\begin{equation}
\Omega \propto \int \left(E-U\right)^{dN/2}\Theta\left(E-U\right) dq^{dN}.
\label{eq:dos}
\end{equation}
This integral has been evaluated analytically for hard spheres, where the system potential energy $U=0$~\cite{Ray1991, Shirts2006, Scalas2015}, but to the best of our knowledge, not for any case of an interaction potential. 

Indeed there may not be an exact analytic solution for the Hamiltonian in Eq.~(\ref{eq:hamiltonian}). Instead we seek an approximate solution, and making the simple observation that the virial theorem holds for these systems, replace $\left(E-U\right)$ with $\left(E-\langle U\rangle_v\right) = \langle K\rangle_v$, where $\langle \dots \rangle_v$ denotes the expected value from the virial theorem. For Eq.~(\ref{eq:hamiltonian}), the virial theorem yields $2\langle K\rangle_v=n\langle U\rangle_v$, and thus 
\begin{equation}
\frac{\langle U \rangle_v}{E} = \frac{2}{n+2}; \hskip1em \frac{\langle K \rangle_v}{E} = \frac{n}{n+2}, 
\end{equation}
with $K$ the system kinetic energy. Thus $\langle K\rangle_v$ can come out of the integral in Eq.~(\ref{eq:dos}), and the integral proceeds as previously described~\cite{Ray1991, Shirts2006,Scalas2015}. 

This substitution cannot be exact: the grain momentum's limit is now set by $ \langle K\rangle_v$, an \emph{average} value, and there are certainly grains with kinetic energy that, at times, are slightly greater than this value. However, we can rely on decreasing fluctuations with increasing $N$, and show that for $N>10$, the number of states beyond this limit is small, and this is a very good approximation.

The resulting pdf of per-grain velocities $v_i$ in 1D is then~\cite{Scalas2015}:
\begin{eqnarray}
\mathrm{pdf}(v_i) 
&=&\mathrm{B}\left(\alpha,\beta,\tilde{v}_i \right)/\left ( 2 \langle v\rangle_v \right), \nonumber \\
&=& \frac{1}{2\langle v\rangle_v} 
\left ( \frac{\Gamma(\alpha+\beta)}{\Gamma(\alpha)\Gamma(\beta)}
\left(\tilde{v}_i\right)^{\alpha-1}\left(1-\tilde{v}_i\right)^{\beta-1} \right ),
\label{eq:pdfv}
\end{eqnarray}
where
\begin{equation}
\tilde{v}_i = \frac{1}{2}\left(1-\frac{v_i}{\langle v\rangle_v}\right),
\end{equation}
with $\langle v\rangle_v^2=2\langle K\rangle_v/m$, and $\alpha=\beta=(N-1)/2$. Also, $\mathrm{B}(\alpha,\beta,\tilde{v}_i)$ is the beta distribution, and $\Gamma$ is the gamma function. In the limit $N\gg1$, Eq.~(\ref{eq:pdfv}) becomes the familiar Maxwell-Boltzmann 1D normal distribution with mean $\mu=0$ and variance $\sigma^2=\langle v\rangle^2_v/N$.

The distribution of kinetic energy per-grain $K_i$ is also given by a beta distribution~\cite{Scalas2015}:
\begin{equation}
\mathrm{pdf}\left(K_i\right) = \mathrm{B}\left(\alpha,\beta; \tilde{K}\right)/\langle K\rangle_v,
\label{eq:pdfki}
\end{equation}
where $\tilde{K}=K_i/\langle K\rangle_v$, $\alpha=1/2$, and $\beta=(N-1)/2$. For $N\gg1$, this becomes the familiar Maxwell-Boltzmann distribution for kinetic energy, a gamma distribution $\mathrm{G}(\alpha,\beta,K_i)$:
\begin{equation}
\mathrm{pdf}\left(K_i\right) = \mathrm{G}(\alpha,\beta,K_i) = \frac{\beta^{\alpha}}{\Gamma(\alpha)}K_i^{\alpha-1}e^{-\beta K_i},
\label{eq:pdfki2}
\end{equation}
where $\alpha=1/2$ and $\beta=N/(2\langle K\rangle_v)$. Interestingly, the possibility of large kinetic energy fluctuations increases with the variance of Eq.~(\ref{eq:pdfki}) (and ~(\ref{eq:pdfki2})), $\langle \delta K_i^2\rangle\equiv\langle K_i^2 \rangle - \langle K_i\rangle^2$;
\begin{eqnarray}
\langle \delta K_i^2 \rangle &=& \frac{2(N-1)}{N^2(N+1)}\left[\left(\frac{n}{n+2}\right)E\right]^2, \nonumber \\ 
&\approx& \frac{2}{N^2}\left [ \left ( \frac{n}{n+2} \right ) E\right ]^2,
\label{eq:rogue}
\end{eqnarray}
which increases to the hard-sphere limit with larger $n$, but rapidly decreases with increasing system size. 

Finally, the distribution of system kinetic energy is given by the Dirichlet distribution~\cite{Scalas2015}, which is a multivariate generalization of the beta distribution and not amenable to visualization or calculation. Alternatively, if we let $K_i$ be independent and identically distributed (i.i.d.) variates drawn from the distributions of either Eq.~(\ref{eq:pdfki}) or (\ref{eq:pdfki2}), then the pdf of $K=\sum_i^N K_i$ can be determined from statistical theory. No such distribution for beta-distributed variates exists for $N>2$~\cite{Nadarajah2015}; however, for the gamma distribution, this is $\mathrm{pdf}\left(K\right) = \mathrm{G}(N/2,N/(2\langle K\rangle_v);K)$.

Although this has the correct mean, comparison with simulation data shows it has the incorrect variance, and after trial-and-error, a better approximation was found to be
\begin{equation}
\mathrm{pdf}\left(K\right) = \mathrm{G}\left(\frac{n+2}{2}\frac{N}{2},\frac{n+2}{2}\frac{N}{2\langle K\rangle_v}; K\right).
\label{eq:pdfk}
\end{equation}
We justify this distribution not only by the excellent empirical match to the distribution calculated from molecular dynamics (MD) simulation, but also from the connection between the variance of system kinetic energy and the specific heat capacity in the microcanonical ensemble. 

In ergodic systems in the thermodynamic limit, Tolman's generalized equipartition theorem~\cite{Tolman1918} applied to Eq.~(\ref{eq:hamiltonian}) yields an average total energy per grain $\left \langle \epsilon \right \rangle = k_B T/2 + k_B T/n$, where $k_B$ is Boltzmann's constant and $T$ is the canonical temperature. The corresponding specific heat per grain is then
\begin{equation}
C_V = \left ( \frac{n+2}{2n}\right ) k_B,
\label{eq:sp}
\end{equation}
which evidently depends \textit{only} upon the exponent in the potential, i.e. there is no grain material, grain size, or temperature dependence.  The equivalence of different statistical ensembles when $N\to\infty$ implies Eq.~(\ref{eq:sp}) is also valid for the microcanonical ensemble in this limit, and when energy is equipartitioned.

It is possible to express the fluctuations in total system kinetic energy in terms of $C_V$ using the approximation found in Refs.~\cite{Lebowitz1967, Rugh1998} which, for 1D systems is 
\begin{equation}
\frac{\langle\delta K^2 \rangle}{\langle K \rangle^2} = \frac{2}{N} \left( 1-\frac{1}{2 C_V}\right ), 
\label{eq:LPV}
\end{equation}
where $C_V$ is in units of $k_B$. Then using Eq.~(\ref{eq:sp}), we have:
\begin{equation}
\langle \delta K^2 \rangle = \frac{2}{N} \left( \frac{2}{n+2}\right) \langle K \rangle ^2, 
\end{equation}
from which the factor of $(n+2)/2$ appears as part of the distribution variance of Eq.~(\ref{eq:pdfk}).

Eq.~(\ref{eq:LPV}) also provides one method to calculate the specific heat per grain from an MD simulation. However, taking an energy derivative of the so-called microcanonical temperature gives the exact formula for the microcanonical specific heat, which in 1D is~\cite{Rugh1998}:
\begin{equation}
C_V = \frac{k_B}{N} \left ( 1 - \frac{(N-4) \langle 1/K^2 \rangle }{(N-2)\langle 1/K \rangle^2 } \right )^{-1}.
\label{eq:Cmc}
\end{equation}
With this equation and Eq.~(\ref{eq:pdfk}), we can compute an approximate $C_V$ for finite microcanonical systems, via analytic approximations  of $\langle 1/K \rangle$ and $\langle 1/K^2 \rangle$.

The cumulative distribution function of $K$ is $F_K(K_0) \equiv P(K< K_0)$. Now consider $X \equiv K^{-\rho}$, where $\rho>0$. By definition $K \ge 0$, thus $F_X(x)=0$ for $x<0$. Meanwhile for $x>0$, $F_X(x) \equiv P(0 < K^{-\rho} \le x) = P(K\ge x^{-1/\rho}) = 1 - P(K < x^{-1/\rho}) =  1- F_K(x^{-1/\rho})$. The $\mathrm{pdf}(X)$ is given by $d F_X(x)/dx$, thus $\mathrm{pdf}(X) = F'_K(x^{-1/\rho})/(\rho x^{(\rho+1)/\rho}) = (\mathrm{pdf}(K)|_{k=x^{-1/\rho}}) / (\rho x^{(\rho+1)/\rho})$. Knowing the pdfs of $1/K$ ($\rho=1$) and $1/K^2$ ($\rho=2$), the means $\langle 1/K \rangle$ and $\langle 1/K^2 \rangle$ can be computed in a standard way. The result is:

\begin{equation}
C_V = k_B \left [ \frac{n+2}{2n} - \frac{1}{N} \left ( \frac{n+2}{n} + \frac{4(N-2)}{nN} \right ) \right ],
\label{eq:Cmc_int}
\end{equation}
which has the form of Eq.~(\ref{eq:sp}) plus an $N$-dependent correction term that vanishes in the thermodynamic limit. Hence Eq.~(\ref{eq:Cmc_int}) provides an estimate for $C_V$ in a large but finite system in which the energy is equipartitioned among the interacting grains.

We point out that all of the distribution functions presented above (per-grain velocity, per-grain kinetic energy, and total system kinetic energy) depend only on the number of grains $N$, the total system energy $E$, and most interestingly, the exponent of the potential energy $n$. To test these distribution functions, we ran MD simulations of a 1D monatomic chain of $N$ grains held between fixed walls and described by the Hamiltonian in Eq.~(\ref{eq:hamiltonian}), which includes grain-wall interactions~\cite{Przedborski2015}. Our grains and walls are steel, and the grains are 6~mm in radius. 

\begin{figure*}[ht]
\centering
\includegraphics[width=0.9\textwidth]{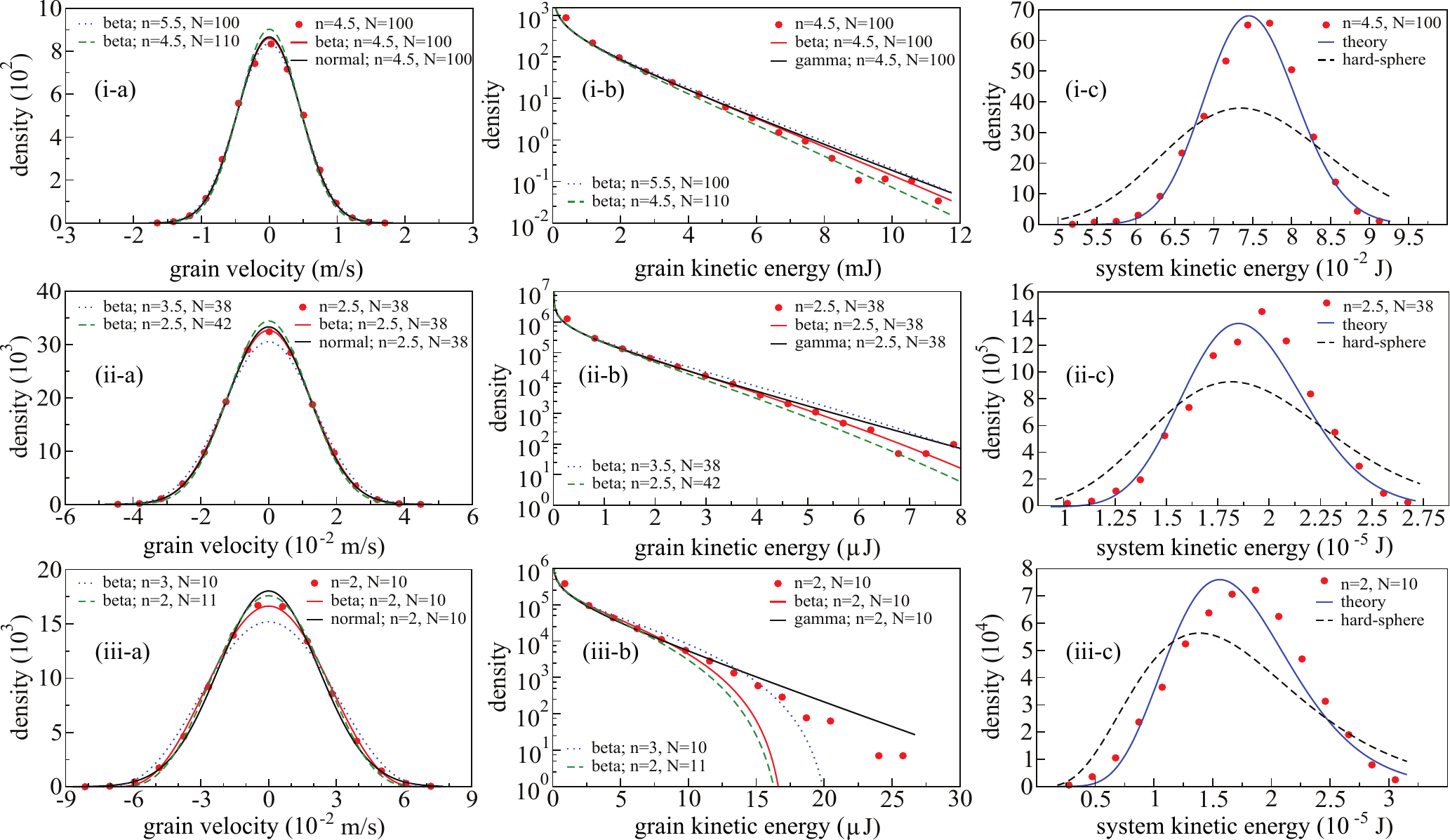}
\caption{(Color online) Distribution of grain velocity, grain kinetic energy, and system kinetic energy for three representative systems.  Results of MD simulations are shown as filled circles. In columns (a) and (b), solid lines are predicted distributions (Eqs.~(\ref{eq:pdfv}),~(\ref{eq:pdfki}),~(\ref{eq:pdfki2})), and dashed/dotted lines are the corresponding distributions with parameters slightly changed to illustrate the sensitivity of Eqs.~(\ref{eq:pdfv}) and~(\ref{eq:pdfki}). In column (c), solid curve is the theoretical prediction Eq.~(\ref{eq:pdfk}), and dashed line is the corresponding hard-sphere distribution.}
\label{fig:dist}
\end{figure*}

We consider values of the potential exponent $n$ from 2 (harmonic) to 5, and system sizes from $N=10$ to 100. A standard velocity Verlet algorithm is used to integrate the equations of motion with a 10~ps timestep, and no dissipation is included.  The grains are set into motion with an initial velocity applied to the first grain only, directed into the chain, causing a SW to propagate through the system. The SW breaks down in collisions with boundaries and in the formation of gaps, creating numerous secondary solitary waves (SSWs). After a period of time, the number of SSWs increases to a point where the system enters into quasi-equilibrium~\cite{Sen2004,Mohan2005,Sen2005,Avalos2007,Avalos2011,Avalos2014}. We allow the system to evolve for a substantial amount of time past this phase change, and at least an order of magnitude longer than previous work has considered. 

The time scale to equilibrium onset is determined by the potential exponent $n$~\cite{Sen2008}, so we adjust the velocity perturbation such that the system arrives at equilibrium quickly. Still, it was necessary to collect at least one second of real time data for $n=2, 2.5, 2.75$, and even longer (up to $6$~s) for larger values of $n$.  Data of grain position and velocity are recorded to file every 1~$\mu$s, though we re-sample the data at time intervals beyond the dampening of velocity autocorrelation (not shown). The deviation from the expected virial $\langle K \rangle_v$ was $<1\%$ for all systems.

In Fig.~\ref{fig:dist} we show the distribution functions obtained from MD simulations and the corresponding expected pdfs (Eqs.~(\ref{eq:pdfv}),~(\ref{eq:pdfki}),~(\ref{eq:pdfki2}), and~(\ref{eq:pdfk})) for three representative systems. In each system, the per-grain velocity data agrees with the beta distribution, Eq.~(\ref{eq:pdfv}), which is nearly identical to the normal distribution for large $N$ (see Figs.~\ref{fig:dist}(i-a), (ii-a)). The difference between the normal and beta distributions becomes apparent for small  systems ($N \lesssim 30$), where the per-grain velocity data fits the beta distribution better.

The grain kinetic energy distributions are presented in Figs.~\ref{fig:dist}(i-b)-(iii-b),  illustrating agreement between MD results and Eq.~(\ref{eq:pdfki}) for large $N$. The difference between Eqs.~(\ref{eq:pdfki}) and~(\ref{eq:pdfki2}) seems pronounced in the log scale with smaller $N$, where the beta distribution has a cutoff before the tail of the MD data. However, for $N=10$, $P(K_i>\langle K \rangle_v)=0.03\%$, while for larger $N$ it's even less. This shows that the limitation of our original virial approximation is quite small. Finally, the sensitivity to $n$ and $N$ are also shown in Fig.~\ref{fig:dist}, with curves of $n+1$ or $1.1N$. They do not agree as well with the data.

Figs.~\ref{fig:dist}(i-c)-(iii-c) contain the distributions of system kinetic energy from MD simulations, along with corresponding Eq.~(\ref{eq:pdfk}), for the three systems. The agreement between MD data and the expected result is very good for $N=100$, see Fig.~\ref{fig:dist}(i-c); less so with decreasing $N$. This is because Eq.~(\ref{eq:pdfk}) develops an increasing skew with decreasing $N$, cf. Figs.~\ref{fig:dist}(i-c) and (iii-c). For comparison, we also present the  distribution without the variance correction, ie. $n=0$, which we call the hard sphere limit, and clearly does not agree with any MD data of interacting grains. 

\begin{figure}[ht]
\includegraphics[width=0.45\textwidth]{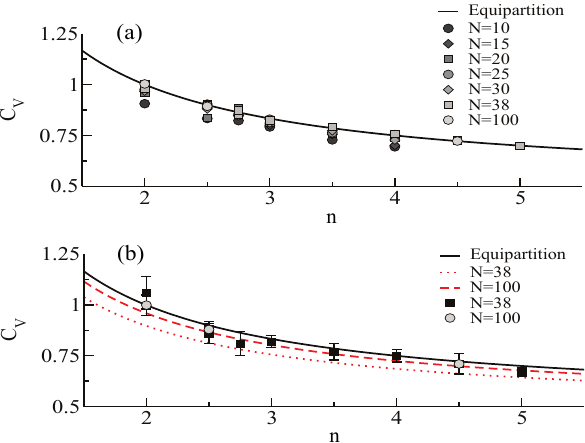}
\caption{(Color online) Specific heat capacities (in units of $k_B$) computed for all MD simulated systems as a function of the exponent on the potential. In (a) we present values obtained from inverting Eq.~(\ref{eq:LPV}), and in (b) values obtained from Eq.~(\ref{eq:Cmc}). The solid line in both plots is the specific heat predicted by the generalized equipartition theorem, Eq.~(\ref{eq:sp}). The dashed lines in (b) are specific heats predicted by Eq.~(\ref{eq:Cmc_int}).}
\label{fig:cv}
\end{figure}

Lastly, we computed the specific heats of MD simulation data using both Eqs.~(\ref{eq:LPV}) and~(\ref{eq:Cmc}). These results are directly compared with $C_V$ predicted by Eq.~(\ref{eq:sp}) shown as the solid line in both Figs.~\ref{fig:cv}(a) and (b), from which it is evident that as $N$ increases, the values calculated by Eq.~(\ref{eq:LPV}) agree very well with the theory. Moreover, even for small ($N\lesssim20$) systems, the deviation from theory is no more than $\sim 10\%$ for Eq.~(\ref{eq:LPV}), and improve with additional statistics. We also present the $n,N$-dependant $C_V$ predicted by Eq.~(\ref{eq:Cmc_int}) as dashed lines in Fig.~\ref{fig:cv}(b), which agrees with the MD data within the error bars for $N=100$. 

The fact that the calculated specific heat agrees with the value predicted by the generalized equipartition theorem for $N\gg1$ provides evidence that energy is indeed equipartitioned in the Hertz chain at late enough times. This finally establishes that the very late-time dynamics of 1D granular chains perturbed at one end with zero dissipation is a true equilibrium phase~\cite{Avalos2011}. The appearance of large fluctuations at late times is thus entirely predictable~\cite{Han2014}. While real granular alignments are inherently dissipative, dissipation-free versions of our systems may be possibly realized as integrated circuits and hence our results may be observable in the laboratory. Finally, quantitative analysis of the QEQ phase may now be possible with this equilibrium theory as the starting point.


These results are also the first empirical demonstration of how the potential energy function can affect the kinetic energy distribution. Shirts et al.~\cite{Shirts2006}, in their calculation of the exact distribution for the finite hard-sphere system, speculate that for attractive potentials $\mathrm{pdf}(K_i)$ would differ somehow, but concede it would be exceedingly complicated to derive. We have shown accurate distributions that may guide attempts to solve Eq.~(\ref{eq:dos}) for finite interaction potentials. 



\begin{acknowledgments}
This work was supported by a Vanier Canada Graduate Scholarship from the Natural Sciences and Engineering Research Council. 
\end{acknowledgments}

\bibliography{Equilibrium}

\end{document}